# High magnetoresistance in graphene nanoribbon heterojunction

S. Bala Kumar[1*], M. B. A. Jalil[1], and S. G. Tan[2]

[1] Department of Electrical and Computer Engineering, National University of Singapore, Singapore, 117576,
[2] Data Storage Institute, (A*STAR) Agency for Science, Technology and Research, DSI Building, 5 Engineering Drive 1, Singapore 117608

[*]Corresponding author. E-mail: brajahari@gmail.com

## Abstract

We show a large magnetoresistance(MR) effect in a graphene heterostructure consisting of an metallic(M) and semiconductor(SC)-type armchair-graphene-nanoribbon(aGNR). In the heterostructure, the transmission across the first subband of the SC-aGNR and M-aGNR is forbidden under zero magnetic-field, due to the orthogonality of the wavefunctions. A finite magnetic-field introduces the quantum hall-like effect, which distorts the wavefunctions. Thus, a finite transmission occurs across the heterojunction, giving rise to a large MR effect. We study the dependence of this MR on temperature and electron energy. Finally, we design a magnetic-field-effect-transistor which yields a MR of close to 100%(85%) at low(room) temperature.

Graphene is a monolayer of carbon atoms arranged in a 2D honeycomb lattice. Due to its lattice structure, graphene possesses many extraordinary properties, which have attracted intense research interests. [1-3] Graphene has also been shown to be a very promising material for future device applications.[4-6] Furthermore, the possibility to pattern graphene into graphene nanoribbons (GNR),[7-9] so as to open up a bandgap via quantum confinement, has increased its versatility as a material for devices. Subsequently, various studies were performed on the transport, magnetic, and optical properties of the GNR.[10-14] More recently, fabrication of GNR with atomically precise edges,[15-17] and thus well-controlled transport properties, has further increased the attractiveness of GNR for practical device application.

GNR can also exhibit magnetoresistive (MR) properties, whereby the resistance of the GNR changes under external magnetic field. The MR effect in graphene-based materials has been investigated in many previous studies, of both experimental[18,19] and theoretical[20-22] nature. In these studies, the contacts are made of ferromagnetic materials, and the MR effect was obtained by changing the relative magnetic orientations of the contacts. Thus in these studies, graphene simply plays a passive role, i.e., as channel material in between the contacts. However, recent experimental study by J. Bai *et. al.*[23] have shown the MR effect in GNR field-effect-transistor (FET) without the inclusion of the ferromagnetic contacts. Similar results were obtained in the theoretical studies,[24,25] in which the MR was ascribed to the B-field induced band gap modification in GNR.

In this paper, we predict a huge MR effect in a dogbone graphene heterostructure, consisting of a metallic-semiconductor aGNR heterostructure (MS-aGNR)[26]. The wavefunction of the first subband of the semiconducting-aGNR (SC-aGNR), and that of the first subband of the metallic-aGNR (M-aGNR) are orthogonal to each other. The zero overlap between the two wavefunctions means that electron transmission is forbidden between the two subbands. Application of external B-field results in quantum Hall-like features (e.g., accumulation of electron density at the edges) and distortion of the wavefunctions. More importantly from the MR point of view, a nonzero overlap occurs between the wave functions, resulting in finite transmission between the two regions. We utilize this property of B-field induced wavefunction distortion to design a magnetic-FET device using the MS-aGNR heterostructure. Based on our transport calculations, we predict an MR of virtually 100% at low temperature, which remains considerable (~ 85%) at room temperature.

We use the nearest neighbor (NN) π-orbital tight binding model[1,27] to calculate the electronic bandstructure of an aGNR under a uniform perpendicular magnetic field (B-field). In the presence of a perpendicular B-field, the magnetic flux passing through each hexagonal ring of the honeycomb carbon structure can be expressed by a dimensionless quantity $\phi = SB_z/\phi_0$, where $\phi_0 = h/e$ is the flux quantum and $S$ is the area of the hexagon. The quantity $\phi$ can be used to characterize the strength of the B-field. The magnetic field of $\vec{B} = (0,0,B)$ induces a vector potential of $\vec{A} = (-By, 0, 0)$, which satisfies $\vec{\nabla} \times \vec{A} = \vec{B}$. Under an applied B-field, following Peierls phase approximation[28], the hopping energy in between neighboring atoms acquires a phase, and the tight-binding Hamiltonian is modified accordingly.[10,12,13,25] We use the NN hopping energy of $t_0$=3eV for the bulk, and 1.12$t_0$ along the edges.[9]

The electron transport behaviors of the aGNR heterostructure are studied using the non-equilibrium Green's function (NEGF)[29] formalism. Based on the NEGF formalism, at temperature $T = 0$ K, the linear response conductance $g$ across the GNR is given by

$$g = g_0 Tr\left[\Gamma_S G^r \Gamma_D \left(G^r\right)^\dagger\right] \qquad (2)$$

where $G^r = [EI - H - \Sigma_S - \Sigma_D]^{-1}$ is the retarded Green's function of the GNR channel, $\Gamma_{S(D)} = i[\Sigma_{S(D)} - \Sigma_{S(D)}^\dagger]$ denotes the coupling between the source (drain) contacts to the GNR, $\Sigma_{S(D)}$ is the self-energy of the source (drain) contacts, and the quantum conductance, $g_0 = e^2/\hbar$.

For the computation of the electron density we consider the following relation:

$$G^n = G^r(f_S \Gamma_S + f_D \Gamma_D) G^{r\dagger} \qquad (3)$$

The electron density at atomic site $i$ is then $D_i = G^n_{ii}/a_0^2$, where $a_0 = 0.142$ nm is the lattice constant. In all our computation, we use aGNRs with the width of around 5 nm, i.e., $N=43$ ($N=41$) for the SC (M)-aGNR, where N is the number of dimer rows.

Let us consider the transport across a aGNR heterojunction. In general, the aGNR can be either metallic (M) or semiconducting (SC), depending on the $N$. When the $N = 3p-1$ ($3p$ or $3p+1$), the aGNR becomes M (SC). In a MS-aGNR heterostructure, formed by connecting M-aGNR and SC-aGNR serially, we would expect the transport gap to be equivalent to the (larger) band gap of the SC-aGNR. However, this is not the case for the two-sided dogbone aGNR-heterostructure, as shown schematically in Fig 1(a). In such a heterostructure, contrary to our expectation, the transport gap is equivalent the difference between the second conduction and second valence subbands.

This may be explained as follows: for GNR, the wave function of states in the first conduction subband-edge is

$$\psi_N(i) = \frac{2}{N+1} \sin\left(\left[\frac{2(N+1)}{3}\right] \frac{i + \lfloor N/2 \rfloor + 1}{N+1} \pi\right) \Theta(|i| - \lfloor N/2 \rfloor), \tag{4}$$

where the row index $i = ..., -1, 0, 1, ...$, $[x]$ is the integer closest to $x$, $\lfloor x \rfloor$ is the closest integer lower than $x$, and $\Theta$ is the unit step-function. Note that, in Eq. 4 the row index is set such i=0 is the middle of the aGNR, and that the $\psi_N(i)$ is finite only for $i = \lfloor N/2 \rfloor - N + 1, ..., \lfloor N/2 \rfloor$. From Eq. 4, it can be shown that

$$\langle \psi_{3p} | \psi_{3p-1} \rangle = \langle \psi_{3p+1} | \psi_{3p-1} \rangle = 0,$$

where $p = 1, 2, 3, ...$, indicating that the first subband of the M and SC aGNRs are orthogonal to each other. We thus expect the electron transmission between these two subbands to be forbidden. This effect is shown in Fig. 1(a), where conductance is suppressed within the first subband of the SC-aGNR.

As shown in Fig. 1(b), when we apply a B-field to the MS-aGNR heterostructure, finite transmission is obtained across the energy range corresponding to the first subband of the SC-aGNR. Fig. 2(a-c) show the spatial variation of electron density with increasing B-field, while Fig. 2(d) and (e) show the modification of the first subband of the SC-aGNR and M-aGNR, respectively, with increasing B-field. From Fig. 2(a-c), one can immediately observe the modification of the spatial distribution of the aGNR's wave function, and hence the electron density (which is proportional to $|\psi|^2$). At zero B-field the electron density is distributed evenly across the transverse direction of the MS-aGNR. However, as the B-field increases, the forward and the backward states are spatially separated to the edges, reminiscent of the edge states of the quantum Hall Effect. As a result, there is enhanced electron density near the top and bottom edges. This wavefunction distortion with applied B-field means that the wavefunctions on the M-aGNR and SC-aGNR regions are no longer orthogonal to one another. As such, $\langle \psi_{SC} | \psi_M \rangle \neq 0$, and thus the MS-junction becomes more transparent. This can be seen from the electron density distribution in Fig. 2(a-c), where at zero B-field the MS-junction is completely reflective, while with increasing B-field, the junction becomes more transparent. These results have significant practical implications, as the conductance of the aGNR is drastically changed upon application of an external B-field, giving rise to a large MR effect.

Next, we analyze the MR effect in the MS-aGNR. The MR ratio is defined as, $\text{MR} = [g(B) - g(0)]/g(B) = 1 - g(0)/g(B)$, where $g(B)$ and $g(0)$ are the conductance of the MS-aGNR at finite and zero B-fields, respectively. Note that in this definition the maximum MR ratio is 100%, corresponding to $g(0)/g(B) \to \infty$. Fig. 3(a) shows the variation of conductance with increasing B-field at different temperature. In general, as expected the conductance increases with increasing B-field, due to the aforementioned distortion of the wavefunction which increases the transparency of the MS-aGNR junction. The corresponding MR as function of B-field is plotted in Fig. 3(b). At low temperature, $\text{MR} \approx 100\%$ for $B = 10$ T, but at higher temperature, MR is suppressed due to Fermi level broadening. We also studied the MR variation at different electron energy. In Fig. 3(c) shows the conductance as a function of electron energy in the absence ($B = 0$) and presence of magnetic field ($B = 10$ T). At low electron energy (below the second conduction band-edge), there is a significant conductance difference between $B = 0$ and $B = 10$ T, with the conductance at zero B-field being two orders of magnitude smaller. However, at electron energy above the second conduction band-edge, the conductance is almost similar for $B = 0$ and $B = 10$ T, and approach $g_0 = h/e^2$ at both B-field values. The convergence in the conductance for $B = 0$ and $B = 10$ above the second conduction band-edge translates into a decrease in MR with increasing electron energy, as shown in Fig. 3(d). The decrease is especially prominent at electron energy close to the second conduction band-edge, since this is the energy threshold at which the MS-aGNR heterojunction becomes transparent even at zero B-field. At higher temperature, the Fermi-broadening effect causes the MR to start decreasing at lower electron energy.

We now propose a magnetic-FET [Fig. 4(a)] similar to that fabricated in Ref. [23], but with the MS-aGNR heterostructure acting as the channel. To model the source-drain metallic contacts, we assume semi-infinite normal metal leads at both ends, whose density-of-state is taken to be constant at 0.03/eV/atom/spin around the Fermi level. In this structure the back-gate is used to vary the Fermi level. The source-drain current across the magnetic-FET is computed as follows

$$I_{SD} = 2/e \int_{-\infty}^{+\infty} g(E) \left[ f_S(E) - f_D(E) \right] dE, \tag{3}$$

where the Fermi distribution function at the source (drain), $f_{S(D)} = \left[ 1 + \exp\left( (E - \mu_{S(D)})/kT \right) \right]^{-1}$, where $\mu_S = E_F + V_{SD}/2$ ($\mu_D = E_F + V_{SD}/2$) is the source (drain) potential, $E_F$ is the Fermi level, and $V_{SD}$ is the applied source drain bias. In all our computations we set the Fermi level at the mid-point between the source and drain bias. Under finite $V_{SD}$, the MR is defined as $\text{MR} = \left[ I_{SD}(B) - I_{SD}(0) \right] / I_{SD}(B)$. The interesting feature of this device is that its conductance can be modulated either by varying the back-gate voltage as in the conventional FETs, or by changing the applied B-field. In general, a larger current is obtained by increasing the B-field. The calculated MR ratio at a magnetic field of $B = 10$ T is shown by the dotted lines in Fig. 4(b-g). Fig. 4(b-d) show that, as in conventional FETs, the current $I_{SD}$ can also be varied by changing the back-gate voltage. The effect of temperature on the conductance and MR can be analyzed by considering Fig. 4(e-f). At the lowest temperature considered ($T = 150$ K), a very large MR ratio of virtually 100% is obtained over a large range of $V_{SD}$, where the transport is primarily occurring within the first subband. With increasing temperature, the effect of Fermi level broadening causes the MR ratio to decrease, as expected. Yet, our results in Fig. 4(f) show that the MR ratio of the device remains high at around 85% even at room temperature. This temperature tolerance may be attributed to the suppression of current over the entire first subband ($\Delta E \sim 0.1$ eV) due to the orthogonality of wavefunctions in the MS-aGNR heterojunction.

In conclusion, we have investigated the MR effect in a MS-aGNR heterostructure. We showed that due to the orthogonality of the wavefunctions across the heterojunction, the transmission between the first subband of the SC-aGNR and the first subband of the M-aGNR is forbidden. However, when a finite B-field is applied, distortion of the wavefunctions removes this orthogonality condition, leading to finite transmission between these subbands. This enables the conductance of the MS-aGNR to be significantly modified by application of B-field, resulting in a large MR effect. By using the MS-aGNR as a channel, we designed a magnetic-FET whose current can by modulated by both E-field and B-field. We showed that the MR of the device is robust to high temperature, and predicted a MR ratio of around 85% even at room temperature. We would like to highlight recent experimental works which demonstrated the fabrication of GNR with atomically precise edges.[15-17] Such precision would enable the electron wavefunctions in actual GNR nanostructures to approach the theoretical predictions of Eq. (4), thus making this study more applicable for future GNR-based device.

We thank Professor Albert Liang for helpful discussions. We also gratefully acknowledge the financial support of SERC Grant No. 092 101 0060 (R-398-000-061-305).

Figure and Caption

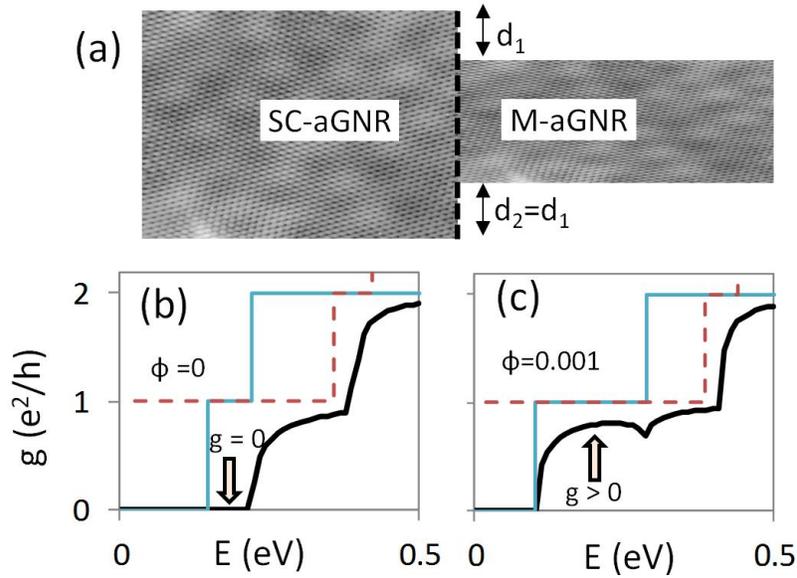

Fig. 1. (a) Schematic diagram of a two-sided dog-bone metallic-semiconducting aGNR (MS-aGNR) heterostructure. (b) Conductance of the MS-aGNR heterostructure at zero B-field. The thin solid (dotted) curve shows the conductance of the ideal SC (M)-aGNR. The conductance of the MS-aGNR heterostructure is suppressed in the energy range corresponding to the first subband of the SC-aGNR. (c) When a finite B-field is applied, however, finite conductance is observed in the first subband. This shows that, within the energy range of the first subband, conductance of the MS-aGNR heterostructure can be modulated by the B-field.

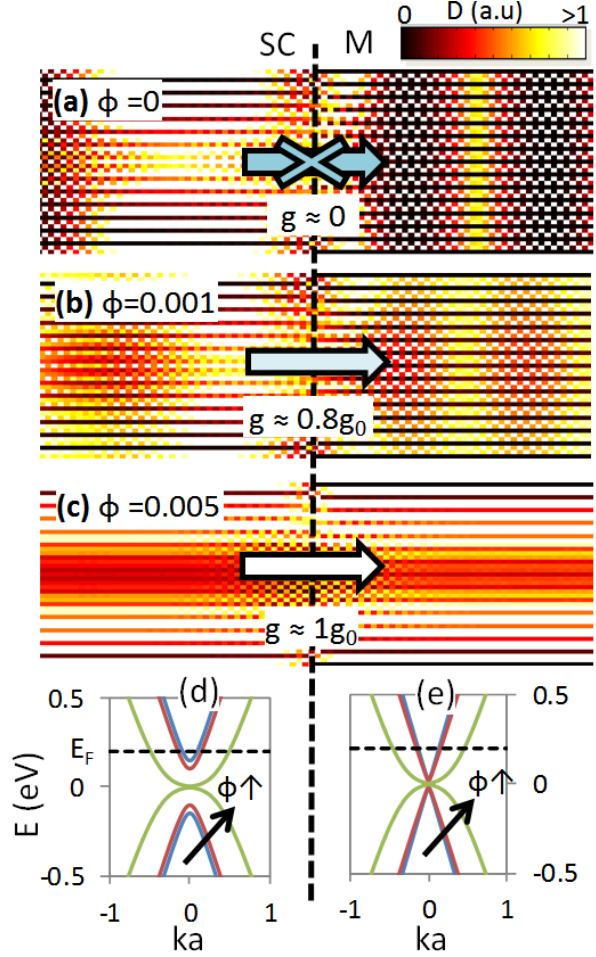

Fig. 2. The spatial distribution of electron density, D at $E_F=0.2$eV (first subband), at B-fields corresponding to flux values of (a) $\phi=0$, (b) $\phi=0.001$, and (c) $\phi=0.005$. The electrons are completely reflected across the MS-junction at $\phi=0$. The MS-junction becomes increasingly transparent with increasing B-field. At higher B-field, the electron density accumulates at the edges. (d) and (e) depict the band structure of the lowest conduction and valence bands at different B-field for (d) SC-aGNR and (e) M-aGNR, and flux values of $\phi=0, 0.001, 0.005$.

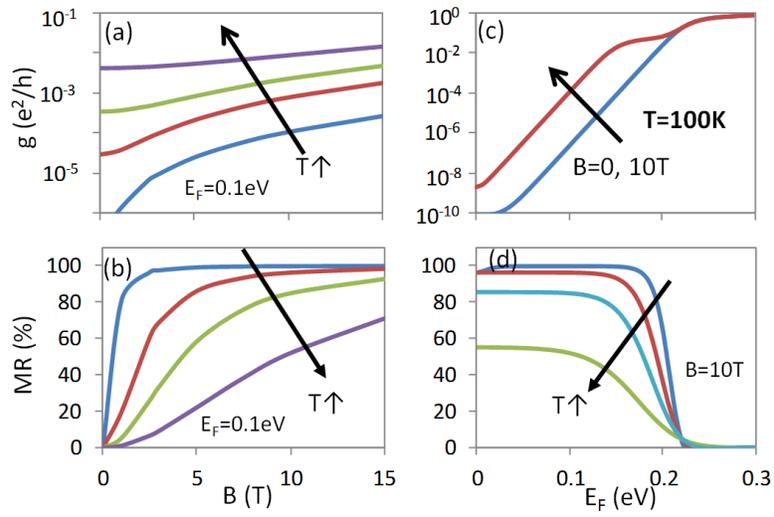

Fig. 3. (a) Conductance and (b) MR ratio, across the MS-aGNR with increasing B-field at different temperatures ($T$ = 100, 150, 200, 300 K). In all the cases, the conductance and MR increases with increasing B-field. (c) Conductance and (d) MR with increasing electron energy at $B$ =10 T. The MR ratio is always suppressed at higher temperature.

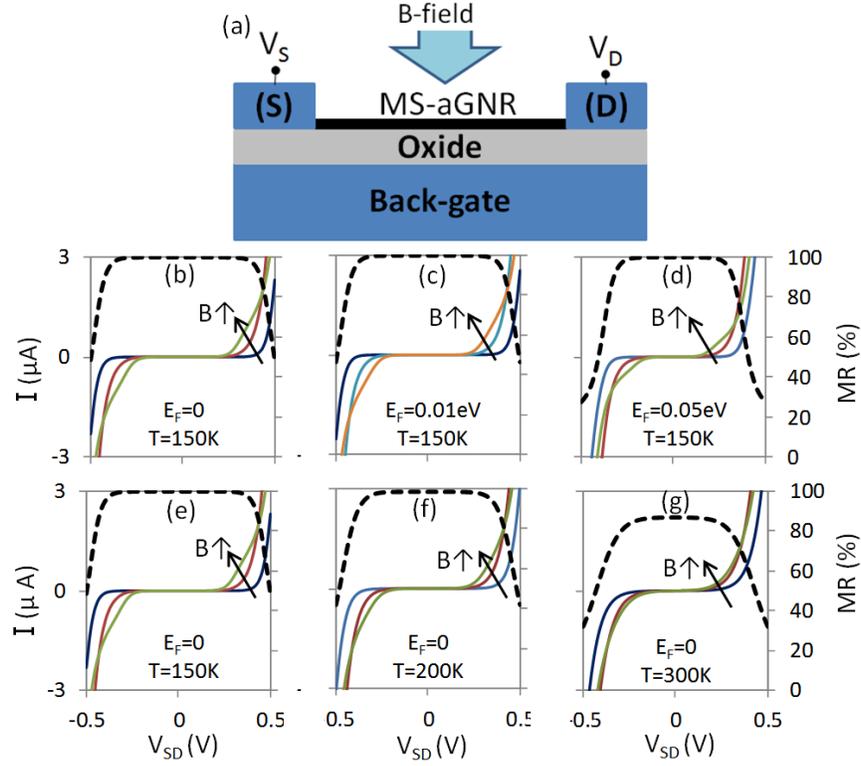

Fig. 4. (a) MS-aGNR magnetic-FET device with metallic contacts and back-gate. The back-gate is used to vary the Fermi energy. $I_{SD}$-$V_{SD}$ curve for (b) $E_F = 0$, (c) $E_F = 0.01$ eV, and (d) $E_F = 0.05$ eV, at low temperature, $T = 150$ K and different B-fields ($B = 0, 5, 10$ T). $I_{SD}$-$V_{SD}$ curve for different temperatures, i.e., (e) $T = 150$ K, (f) $T = 200$ K, and (g) $T = 300$ K, at electron energy $E_F = 0$ at different B-field. In all the plots, the dotted lines represent the MR at $B = 10$ T.

**References**

1. A. H. Castro Neto, F. Guinea, N. M. R. Peres, K. S. Novoselov, and A. K. Geim, Reviews of Modern Physics **81,** 109 (2009).
2. A. K. Geim, Science **324,** 1530 (2009).
3. A. K. Geim and K. S. Novoselov, Nature Materials **6,** 183 (2007).
4. C. Berger, Z. Song, T. Li, X. Li, A. Y. Ogbazghi, R. Feng, Z. Dai, A. N. Marchenkov, E. H. Conrad, P. N. First, and W. A. de Heer, The Journal of Physical Chemistry B **108,** 19912 (2004).
5. N. S. Norberg, G. M. Dalpian, J. R. Chelikowsky, and D. R. Gamelin, Nano Letters **6,** 2887 (2006).
6. K. S. Novoselov, D. Jiang, F. Schedin, T. J. Booth, V. V. Khotkevich, S. V. Morozov, and A. K. Geim, Proceedings of the National Academy of Sciences of the United States of America **102,** 10451 (2005).
7. L. Brey and H. A. Fertig, Physical Review B **73** (2006).
8. K. Nakada, M. Fujita, G. Dresselhaus, and M. S. Dresselhaus, Physical Review B **54,** 17954 (1996).
9. Y.-W. Son, M. L. Cohen, and S. G. Louie, Physical Review Letters **97,** 216803 (2006).
10. Y. C. Huang, C. P. Chang, and M. F. Lin, Nanotechnology **18,** 495401 (2007).
11. K. Tada and K. Watanabe, Physical Review Letters **88,** 127601 (2002).
12. K. Wakabayashi, Physical Review B **64,** 125428 (2001).
13. K. Wakabayashi, M. Fujita, H. Ajiki, and M. Sigrist, Physical Review B **59,** 8271 (1999).
14. K. Wakabayashi and M. Sigrist, Physical Review Letters **84,** 3390 (2000).
15. J. Cai, P. Ruffieux, R. Jaafar, M. Bieri, T. Braun, S. Blankenburg, M. Muoth, A. P. Seitsonen, M. Saleh, X. Feng, K. Mullen, and R. Fasel, Nature **466,** 470 (2010).
16. P. Ruffieux, J. Cai, N. C. Plumb, L. Patthey, D. Prezzi, A. Ferretti, E. Molinari, X. Feng, K. Müllen, C. A. Pignedoli, and R. Fasel, ACS Nano (2012).
17. S. Blankenburg, J. Cai, P. Ruffieux, R. Jaafar, D. Passerone, X. Feng, K. Müllen, R. Fasel, and C. A. Pignedoli, ACS Nano **6,** 2020 (2012).
18. E. W. Hill, A. K. Geim, K. Novoselov, F. Schedin, and P. Blake, IEEE Trans. Magn. **42,** 2694 (2006).
19. W. H. Wang, K. Pi, Y. Li, Y. F. Chiang, P. Wei, J. Shi, and R. K. Kawakami, Physical Review B **77,** 020402 (2008).
20. L. Brey and H. A. Fertig, Physical Review B **76,** 205435 (2007).
21. K.-H. Ding, Z.-G. Zhu, and J. Berakdar, Physical Review B **79,** 045405 (2009).
22. W. Y. Kim and K. S. Kim, Nat Nano **3,** 408 (2008).
23. J. Bai, R. Cheng, F. Xiu, L. Liao, M. Wang, A. Shailos, K. L. Wang, Y. Huang, and X. Duan, Nat Nano **5,** 655 (2010).
24. S. Bala kumar and J. Guo, Nanoscale **4,** 982 (2012).
25. S. B. Kumar, M. B. A. Jalil, S. G. Tan, and G. C. Liang, Journal of Applied Physics **108,** 033709 (2010).
26. S. Hong, Y. Yoon, and J. Guo, Applied Physics Letters **92,** 083107 (2008).
27. R. Saito, M. Fujita, G. Dresselhaus, and M. S. Dresselhaus, Appl. Phys. Lett. **60,** 2204 (1992).
28. R. E. Peierls, Z. Phys. **80,** 763 (1933).
29. S. Datta, *Quantum Transport: Atom to Transistor* (Cambridge University Press, Cambridge, 2005).